# Title: The Unresolved Fine Structure Resolved - IRIS observations of the Solar Transition Region

**Authors:** V. Hansteen[1*], B. De Pontieu[1,2], M. Carlsson[1], J. Lemen[2], A. Title[2], P. Boerner[2], N. Hurlburt[2], T.D. Tarbell[2], J.P. Wuelser[2], T.M.D. Pereira[1], E.E. De Luca[3], L. Golub[3], S. McKillop[3], K. Reeves[3], S. Saar[3], P. Testa[3], H. Tian[3], C. Kankelborg[4], S. Jaeggli[4], L. Kleint[5,2], J. Martínez-Sykora[5,2]

**Affiliations:**

[1] Institute of Theoretical Astrophysics, University of Oslo, Post Office Box 1029, Blindern, NO-0315, Oslo, Norway.

[2] Lockheed Martin Solar and Astrophysics Laboratory, 3251 Hanover st., Org. A021S, Bldg.252, Palo Alto, CA, 94304, USA.

[3]Harvard-Smithsonian Center for Astrophysics, 60 Garden st., Cambridge, MA 02138, USA.

[4]Department of Physics, Montana State University, Bozeman, P.O. Box 173840, Bozeman MT 59717, USA.

[5] Bay Area Environmental Research Institute, 596 1st St West, Sonoma, CA, 95476 USA.

*Correspondence to: viggoh@astro.uio.no

**Abstract**: The heating of the outer solar atmospheric layers, i.e., the transition region and corona, to high temperatures is a long standing problem in solar (and stellar) physics. Solutions have been hampered by an incomplete understanding of the magnetically controlled structure of these regions. The high spatial and temporal resolution observations with the Interface Region Imaging Spectrograph (IRIS) at the solar limb reveal a plethora of short, low lying loops or loop segments at transition-region temperatures that vary rapidly, on the timescales of minutes. We argue that the existence of these loops solves a long standing observational mystery. At the same time, based on comparison with numerical models, this detection sheds light on a critical piece of the coronal heating puzzle.

**One Sentence Summary:** IRIS observations prove that a large fraction of the solar transition region emission is due to low lying, relatively cool, loops having no thermal contact with the corona.

**Main Text:** The outer solar atmosphere between the $10^4$K chromosphere and the $10^6$ K corona, the so-called transition region, has long puzzled solar physicists *(1).* It has been difficult to reconcile measured intensities and motions, either directly observed or inferred from spectra, with models of the energy and mass exchange between the cooler chromosphere and hot corona. For one, the observed intensities of lower transition region lines are much greater than can be accounted for by thermal conductive flux flowing back from the corona. Furthermore, lower transition region lines show, on average, significant Doppler redshifts of order 10 km/s *(1),* one third the speed of sound. Based on indirect spectroscopic evidence from High Resolution

Telescope Spectrograph (HRTS) and Skylab spectra it was already postulated in 1983 that the dominant emission from lines formed in the transition region occurs in structures magnetically isolated from the corona called the "unresolved fine structure" (UFS) *(2,3,4,5)*. However, the following decades have not brought consensus that the UFS has been directly observed, nor indeed that it contributes significantly to transition region emission *(6,7,8,9,10)*. As a result, our understanding of coronal heating has not advanced substantially.

We exploit the high spatial and temporal resolution of the recently launched IRIS satellite to reveal structures remarkably similar to those postulated to comprise the UFS. Images of the lower transition region at the solar limb with the IRIS slit jaw camera (11) in the Si IV 1400 Å filter or in the C II 1330 Å filter invariably show bright low lying loops or loop segments in quiet sun regions (Fig. 1, Movies S1 and S2). In addition to these bright structures, a much fainter component forms a background that extends up to 10 arcsec above the limb. The background includes a large number of linear structures, with properties similar to the well known spicules observed from the ground in the Hα 656.3 nm line. Here we concentrate on the brighter loop-shaped objects that appear to be magnetically isolated from the corona and are at transition region temperatures.

Viewing the same limb with the Solar Dynamics Observatory Atmospheric Imaging Assembly (SDO/AIA) instrument *(12)* in the 304 Å channel (which is dominated by He II, at 100,000 K) and the coronal 171 Å (Fe IX/X at ~$10^6$), and 193 Å (Fe XII at ~1.5 $10^6$ K) channels *(13),* we do not clearly observe UFS-related structures (Fig. 1, Movie S3). This is for two reasons: 1. the AIA spatial resolution is insufficient to resolve the structures discussed here, and 2. bound-free absorption by neutral hydrogen and singly ionized helium renders the AIA opaque to extreme ultraviolet emission, thereby shielding the lower transition region and making the UFS nearly invisible.

One of the most striking features of UFS loops is temporal variability. Isolated UFS loops light up, either partially or wholly, and show large changes on scales down to the shortest cadence data inspected so far (every 4 s), examples of which are shown in region of interest (ROI) 2 and ROI 3 of Fig. 1. On the other hand, a system of loops, in which individual loops vary from exposure to exposure, can remain recognizable as a system over periods extending to several tens of minutes. The temporal evolution of the three regions of interest are shown in Fig. 1 (more loops are displayed in Fig. S1). ROI 1 was observed with a cadence of 54 s and is an example of a fairly long lived "nest" of loops that remains active during the entire 40 minute span the observations lasted. Movies S1, S2, and S3 show these nests to be comprised of many loops with more or less co-spatial foot points that light up and darken episodically.

Fully formed loops are seldom seen. Rather, loops appear to be lit up in segments, with each segment only being visible for roughly a minute. There is also a tendency for the UFS loops to appear to rise with time, as can be seen in ROI 2 (Fig. 1).

We find that the UFS loops have a full length of 4 -- 12 Mm ($10^6$ meters), a maximum height of 1 -- 4.5 Mm with an average of 2.5 Mm, and a median intensity of 40-50 DN/s (Figs. 2, *14)*. This intensity is larger than the measured intensity of the background "spicules" – longer nearly radial features - of 15 DN/s, which is also apparent from visual inspection of Fig. 1. While the detailed filling factor of either component is not well known (given the superposition at the limb and their limited visibility on the disk), it is clear that both resolved components have a significant role in the lower transition region emission, with the relative contribution dependent on the local magnetic field topology.

By aligning the slit along the limb, IRIS also allows one to gather spectral data of the UFS (example in Fig. 3). The spectrum shows large excursions as a function of position along the loop, implying large plasma velocities, towards the red as well as towards the blue. We find extreme line profiles at the upper loop foot point – the portion of the loop which meets the underlying atmosphere - during the entire 200 s lifetime of the UFS loop, with red-ward excursions of 70-80 km/s. This is two to three times the speed of sound in a 80,000-100,000 K plasma. We observed the spectral properties of several UFS loops and find that such high velocities occur often, though not always. This indicates that the UFS loops are locations of episodic and violent heating.

How can we put these observations into the context of coronal heating models and the structure of the upper solar atmosphere? Models that assume most transition region emission stems from loops connected to the corona, and therefore whose temperature structure is maintained by thermal conduction, lead to predicted intensities much smaller than those observed. Alternative models proposed the existence of low-lying cool loops *(15,16, 17)*, but these static models cannot be reconciled with the highly dynamic loops we observe here. Guidance comes from 3D modeling: low-lying episodically heated loops that seldom or never reach coronal temperatures naturally arise in 3D models and predict *(18,19, see also 20)* highly dynamic spectral lines originating in the lower transition region. Thus, several properties of UFS loops are remarkably similar to those found in recent realistic 3D models spanning the convection zone to the corona.

The typical loop height of the brightest cool loops that spontaneously arise in 3D simulations (Fig. 4) is less than 4 Mm above the photosphere. The lifetime of individual "strands", or loops, that maintain plasma in the temperature range required for emission in the Si IV line is a few hundred seconds or less. Both of these properties are very similar to what is observed. However, Doppler velocities measured at the origin of the line profile show line-of-sight velocities on the order of 20-30 km/s, which is less than that reported for our observations.

Why do these loops form? Heating in the upper chromosphere and corona will proceed along and be guided by the loop magnetic field. This could occur either through the dissipation of waves or through the dissipation of stresses built up as a result of foot point motions *(21).* While long loops lose energy through thermal conduction, shorter loops are denser at their apices and will therefore lose energy efficiently through radiative losses, which scale as the density squared *(e.g., 18,19,20).* The cooling time of low-lying loops is short (a few minutes or less) and a reduction in the heating rate ensures rapid cooling. At heights of 5 Mm or less above the photosphere, these models therefore predict short low-lying loops that are episodically heated to ~500,000 K or less. The loops cool rapidly thereafter rather than being heated to coronal temperatures.

The discrepancy between observed and modeled velocities could be an indication that the models correctly predict the spatial distribution and episodic nature of the heating in the corona, but the detailed nature of the heating mechanism may not be exactly reproduced. The spatial distribution of loops is largely independent of the heating mechanism, and is instead  set by the structure of the field. We thus expect low-lying loops to occur in any realistic 3D model. However, the temporal properties of the loop emission or Doppler velocities are a direct result of the heating process, and comparison of synthesized and observed data could validate a given model.

What sets the low height of the observed loops? It is likely that these dynamic small-scale loops in the low solar atmosphere are associated with ubiquitous weak magnetic field in the solar photosphere *(22,23,24).* These continuously emerging weak fields occur on granular scales with opposite polarities separated by a few Mm. Our observations of the low loop heights and the sometimes persistent "nests" of loops are fully compatible with theoretical predictions. If such weak fields are present, they should form a multitude of dynamic low-lying loops *(16),* especially when they interact with the strong magnetic field in the quiet sun network *(25).* The apparent rise of some of the observed loops is likely caused by the emergence of the loops into the atmosphere *(26).*

Based on their properties, and in comparison with 3D models, we conclude that the short low-lying loops observed with IRIS constitute a set of low-lying magnetic structures whose plasma is episodically heated. The thermal properties of these loops are determined by their high density, which causes efficient radiative cooling and thus prevents the occurrence of coronal temperatures. Higher-lying loops are presumably heated in much the same way, but with radiative losses are much less efficient at lower densities, and temperatures must rise to > 1 MK in order to balance heating with losses due to thermal conduction. By revealing the existence and properties of these previously unresolved fine-structured loops, we have obtained direct insight into the otherwise difficult to observe coronal heating mechanism. The episodic nature, height distribution and high velocities of UFS loops provide strict constraints on recent 3D models of the coronal heating problem. Further observations and more advanced modeling of these loops are critical for determining the relation between the topology of the magnetic field in the photosphere and vigorous heating in the outer atmosphere.

**Acknowledgments:** IRIS is a NASA Small Explorer mission developed and operated by LMSAL with mission operations executed at NASA Ames Research center and major contributions to downlink communications funded by the Norwegian Space Center (NSC, Norway) through an ESA PRODEX contract. This research was supported by the Research Council of Norway through the grant “Solar Atmospheric Modelling” and through grants of computing time from the Programme for Supercomputing. This work is supported by NASA contract NNG09FA40C (IRIS), the Lockheed Martin Independent Research Program and European Research Council grant agreement No. 291058. SAO authors are supported by LMSAL contract 8100002705. The data presented in this paper may be found on the Hinode Science Data Centre Europe (www.sdc.uio.no).

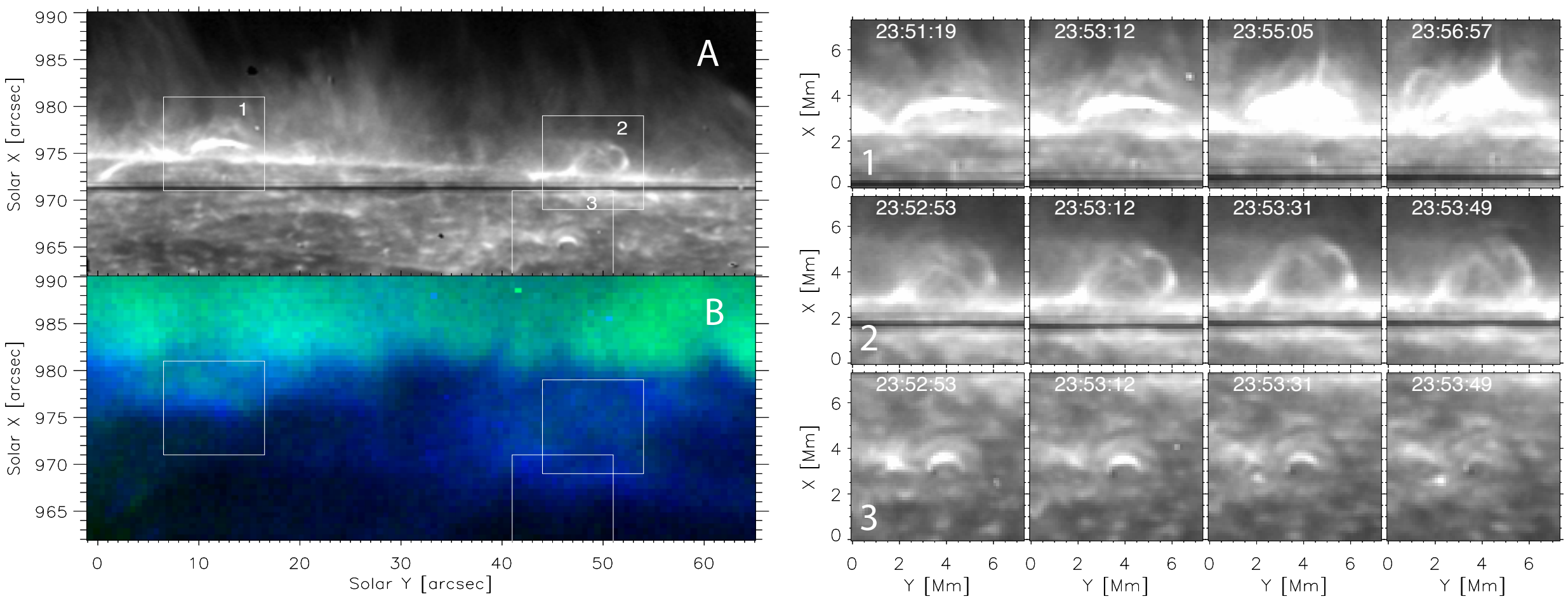

**Fig. 1. IRIS Si IV 1400 Å slit jaw images reveal highly dynamic, low-lying loops at transition region temperatures.** (A)UFS-loops on the western solar limb. The slit is evident as a dark line near solar-x=971 arcsec. (B) the same field of view is shown, but with SDO/AIA images: the coronal 171 Å (blue) and 193 Å (green) filters (Movie S3 shows the same field of view, but with the AIA He II 304 Å filter). The rapid evolution of the UFS in three regions of interest (ROI) is demonstrated in the small panels. The time of each exposure (hh:mm:ss) is indicated at the top of the panels.

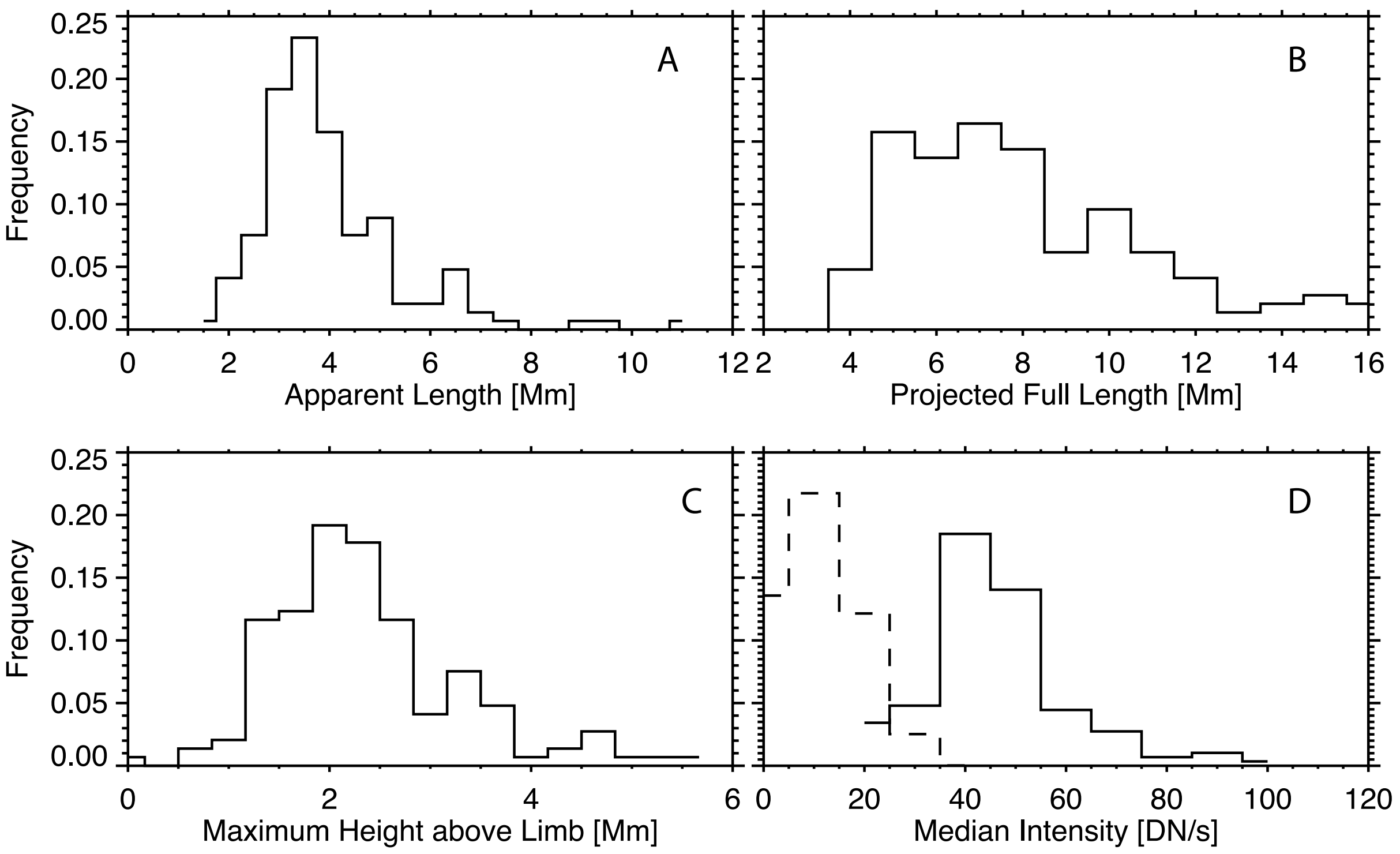

**Fig. 2. UFS loops are short, bright and low-lying**. Properties of 85 UFS loops taken from a limb observation data set: (A) apparent length of the visible loop segments, (B) projected full length along loop, (C) maximum height, and (D) median intensity of the visible loop segments. The intensity measured for the dimmer "spicular" background is shown with a dashed line.

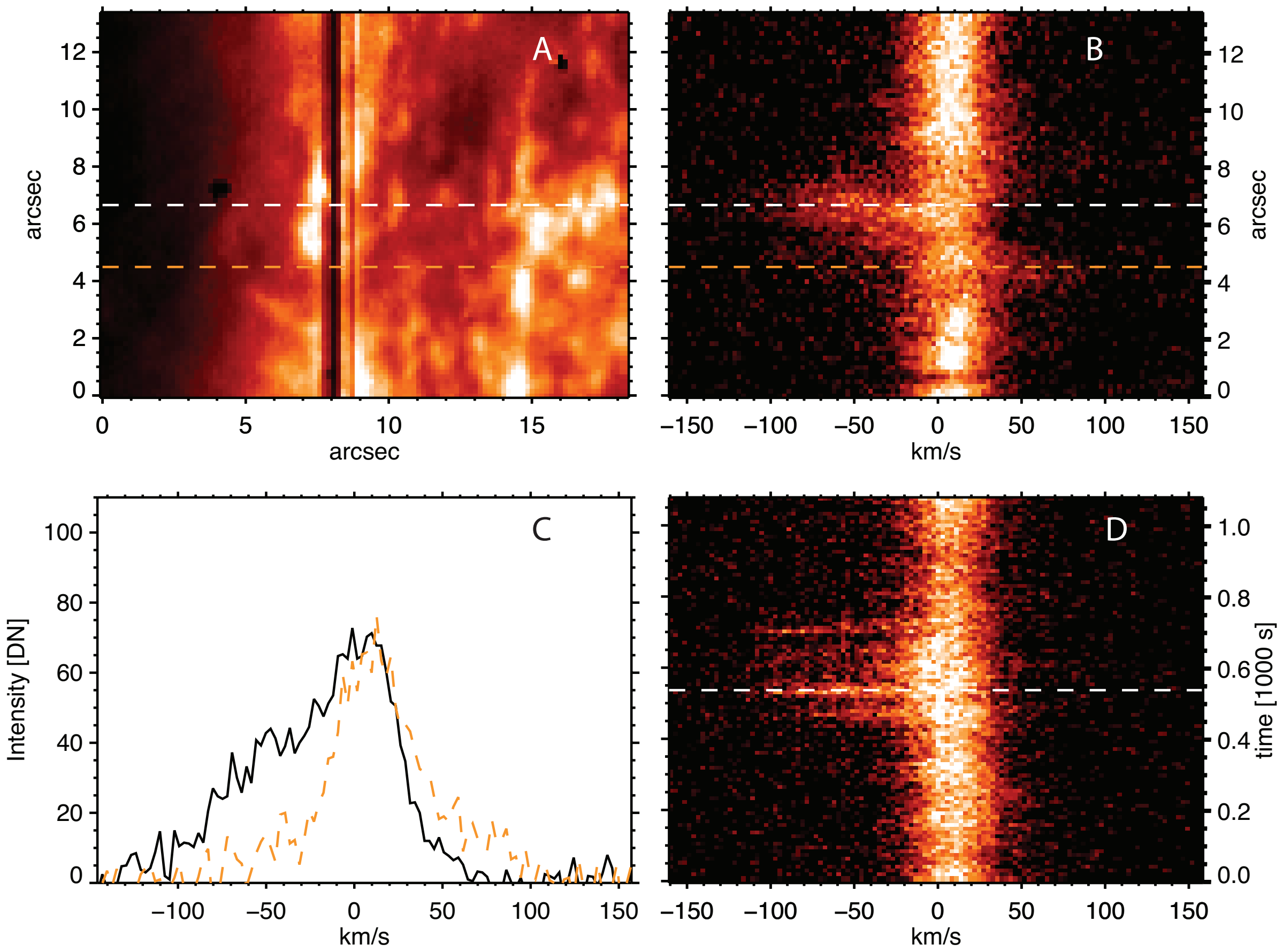

**Fig. 3. UFS loops display large, rapidly evolving Doppler shifts and non-thermal velocities.** (A) The slit position along the limb (SJI 1400 Å). (B) The corresponding spectral scans of the Si IV 1393 Å line are shown as function of position along the slit and as a function of time at the location of the UFS loop foot point (D) located at 6.6 arcsec. In panels B and D, the dashed white and orange lines show the locations of the foot points. (C) The corresponding line profiles at these foot point locations (540 s) are shown in solid and gold dashed curves respectively.

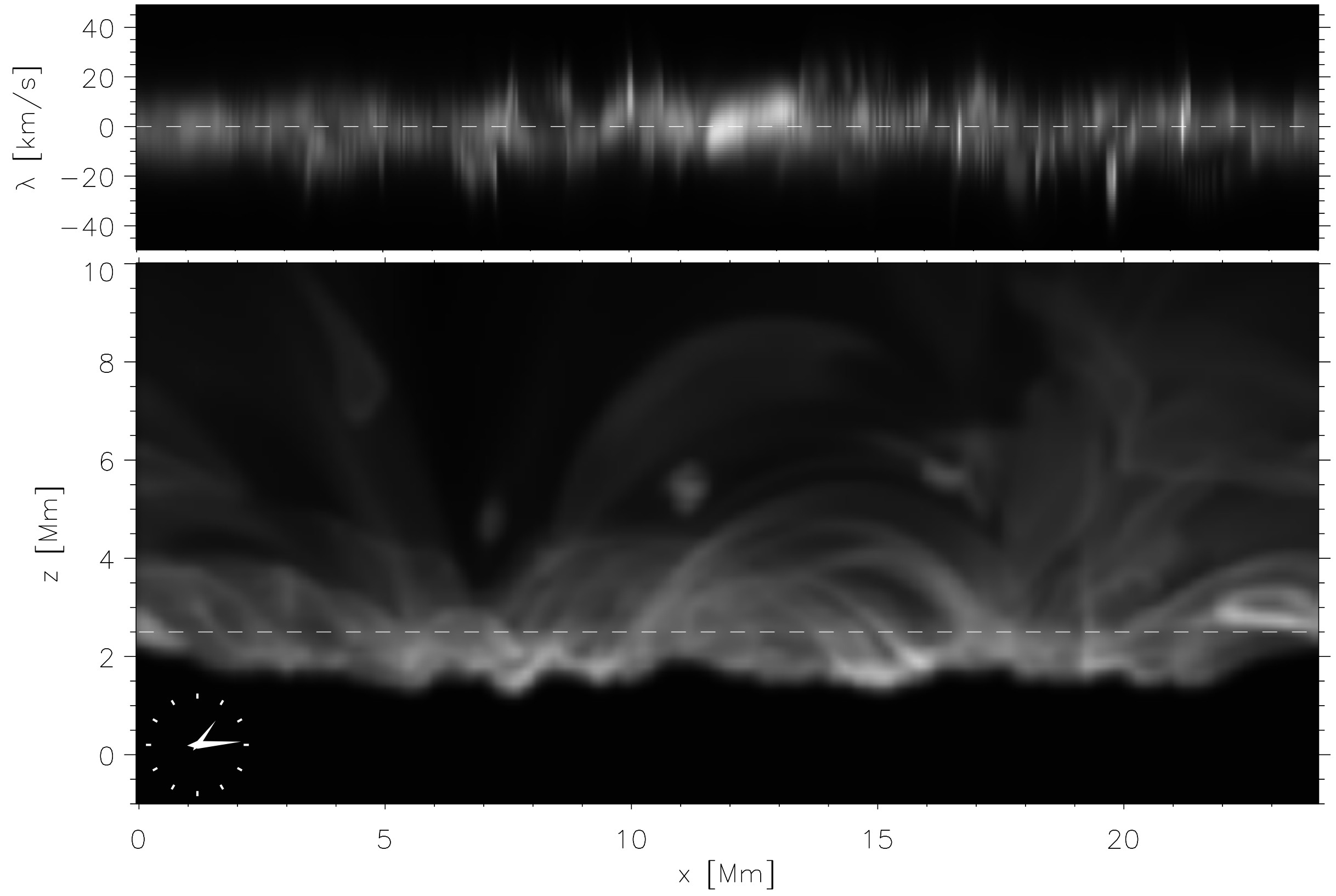

**Fig. 4. Episodic structures with properties similar to UFS loops arise in advanced 3D numerical simulations.** The synthetic line profile of the Si IV 1393 Å line (top) and the corresponding synthetic intensity images of the solar limb (bottom) are shown. The dashed white line in the lower panel shows the location sampled for the synthetic spectrum. The resolution of the intensity images has been degraded to the IRIS spatial resolution of 0.33 arcsec. Movie S4 shows the temporal evolution of this simulation.

**Supplementary Materials:**

Materials and Methods

Discussion

Table S1

Movies S1-S6

References (*27-32*)

## Materials and Methods

Observations

IRIS Observations

To study the UFS we have used several different IRIS datasets, listed in table S1. All of the IRIS data were calibrated to level 2, i.e., including dark current, flat-field and geometric correction (*11*). The raster scans are created by scanning the IRIS slit across the solar disk at either dense (0.35"), or coarse (2") steps. The slit-jaw images (SJI) were corrected for dark-current and flat-field, as well as internal co-alignment drifts.

IRIS Passbands

In the paper we use slit-jaw images in the 1400 Å passband. These contain emission from the Si IV 1394/1403Å lines which are formed in the transition region and according to the CHIANTI database (*27*), at a temperature of ~80,000 K when formation occurs in equilibrium. We see essentially the same UFS structures in the 1330 Å passband, which is dominated by the C II 1334/1336 lines formed in the upper chromosphere/lower transition region, at ~20,000 K. Both FUV SJI channels have a broad passband that also includes continuum formed in the upper photosphere/lower chromosphere. Off-limb the emission in the 1400 Å channel stems essentially only from transition region emission in the Si IV lines. While there are considerable quantities of cool material just above the limb, as evidenced by the lower panel of Figure 1 which shows a dark band of opaque material stretching some 5 arcsec above the 1400 Å passband limb. AIA continuum opacities are dominated by H and He and material cool enough that these elements are not ionized ($T < 20\ 10^4$ K) will be very opaque. In the IRIS passband the continuum opacity is due to neutral Si which is much less abundant than H or He: There is simply not enough neutral Si just above the limb to give significant opacity in the 1400 Å passband.

The slit-jaw images in the 1400 Å passband contain a number of lines (as well as continuum emission when observing the solar disk). An absolute calibration has been reported in the IRIS instrument paper (*11*) and effective area curves can be found in the IRIS software package. With 12 electrons per DN, a pixel size of $6.5\ 10^{-13}$ sr, an effective area of 0.6 $cm^2$, 2 electrons per EUV photon, and an energy per photon of $1.4\ 10^{-11}$erg at 1400 Å, we find 215 erg/s/$cm^2$/sr per DN. Based on observed spectra we can then estimate that the dominant Si IV 1393 Å line represents 60-70% of the total intensity in the slit jaw images above the limb, and the Si IV 1403 Å line most of the rest. Thus, an observed DN of 40 gives a total Si IV 1393 Å intensity of roughly $5\ 10^3$ erg/s/$cm^2$/sr.

The IRIS spectra of the Si IV 1393 Å line shown in Figure 3 is sampled at 12.72 mÅ (2.7 km/s) in the FUV2 passband (1389-1407 Å).

AIA Observations

The AIA data (*12*) acquired for this paper covers the period 9-Dec-2013 23:30 to 10-Dec-2013 00:15, and consists of level 1.5 data which is flat field and dark current corrected as well having the various filters spatially co-aligned and put on an absolute solar XY scale. The IRIS - AIA co-alignment was performed by comparing the IRIS SJI 1400 Å channel with the AIA 1700 Å channel for similar structures and thereafter ensuring that the AIA internal co-alignment was consistent, by comparing co-temporal AIA images in the passbands considered. We estimate the accuracy of the co-alignment to be of order 1-2 arcsec.

The three AIA channels used here, 304 Å, 171 Å and 193 Å, are dominated by He II 304 Å (100,000 K), Fe IX/X ($10^6$ K), and Fe XII (1.5 $10^6$ K), respectively. There is a certain contribution from cooler, transition region, lines in the coronal channels and from a hot Si XI line in the 304 Å channel (see *13*), but any eventual "contamination" is not important for the results of this paper.

Methods

Loop properties histograms

In order to quantify the properties of the UFS we have measured the properties of 85 loops taken from a limb observation data set on 2-Oct-2013 at 07:40 UT. We have measured the apparent length and intensity of the visible loop segments and thereafter fit each segment with a (fourth order) polynomial extended to the limb, and from this fit measured the projected full length and maximum height above the limb of the loops.

Numerical Model

The simulation behind Figure 4 is based on a 'realistic' 3D numerical experiment using the Bifrost code (*28*). 'Realistic' in this context means that the numerical experiments are meant to produce synthetic observables that can be compared directly with observations. In this model, convection is driven by optically thick radiative transfer and losses from the photosphere and lower chromosphere, including scattering. An effective temperature close to $T_{eff}$=5780 K is maintained by setting the entropy of inflowing material entering the bottom boundary. Also, in the middle/upper chromosphere, the transition region, and the corona, radiative losses are parameterized using recipes derived by Carlsson & Leenaarts (*29*). Field aligned conduction is included. To maintain a reasonable time step, the conduction operator is solved using an implicit multi-grid method. Non-equilibrium hydrogen ionization is included as described in (*28*). The model is run with an average unsigned magnetic field strength of 50 G with two dominant polarity regions ~8 Mm apart in a box that spans 24x24x17 $Mm^3$ on a grid with 48 km horizontal resolution and 19-100 km vertical resolution. After an initial period of about 30 minutes of solar time, the convective motions stress the field sufficiently to support coronal heating, maintaining a temperature of > $10^6$ K. Movie S4 shows the simulation starting at roughly 1 hour after the introduction of the magnetic field. Figure 4 is taken from a snapshot at 1 hour 22 minutes. While the model resolution is sufficient to resolve the low temperature gradient loops that dominate lower transition region emission, one could speculate that increasing the resolution will lead to smaller, more violent, individual heating events along the loop axes in the models and thereby to more extreme Doppler profiles in the modeled UFS loops.

The simulation is publicly available as part of the IRIS data product at www.sdc.uio.no.

**Discussion**

Comparison with previous observations

The concept of unresolved fine structure (UFS) was originally surmised from low resolution spectroscopic evidence (*2-6*). As higher resolution instruments have become available, various studies have claimed to have resolved the UFS. We argue below that the very small-scale (lengths of 5 arcsec and widths less than 1 arcsec) and the highly dynamic nature (lifetime of less than one minute) of the loops we report on in this paper have kept these features out of reach of previous instrumentation. This is best illustrated by movies S5 and S6 in which UFS loops can be solidly traced in time and space in the 20 s cadence SJI movies (top panels), but not reliably in the high resolution (0.33 arcsec) IRIS 2.5 min cadence spectroheliograms because of the very fast temporal evolution of the loops. This issue becomes more dramatic for lower-resolution spectrographs such as SUMER which has a spatial resolution of 2 arcsec and a temporal resolution similar or worse than IRIS.

For example, Feldman et al. (*4*) analyzed a 'typical quiet sun' region in a SUMER raster image of the entire solar disk in the C III 977 Å line with a pixel size of 1 arcsec and a raster step size of 1.14 arcsec, each raster position having an exposure time of 7 s. Clumps of emission that 'appear to follow the quiet sun network' are reported to consist of densely packed 'bright looplike structures' that on average are 10-20 arcsec long and with widths below the SUMER resolution. Further, cell interior looplike structures are reported to have a wider distribution of lengths (10-50 arcsec). The CIII 977 Å line is formed at a temperature very similar to the Si IV 1393 Å line (80,000 K). The lengths found in this study are significantly shorter than those found by Feldman et al. This discrepancy is a consequence of the SUMER 2 x 2 arcsec$^2$ spatial resolution, which means that a typical IRIS UFS loop would be spread over 2 or 3 resolving elements, so that identification of a "loop" becomes problematic. In addition, and perhaps more importantly, assembling a rastered image takes longer than the typical timescale of a UFS loop. In other words, the UFS loops we report on here are so dynamic and small-scale they were not identifiable as loops with a typical SUMER raster. We also note that in images taken on the disk it is extremely difficult to separate spicular structures and true low-lying cool loops, as discussed below.

Patsourakos et al (*10*) utilized VAULT spectroheliograph observations centered on the Lyman α line with a 70 Å bandpass filter line taken with the VAULT instrument. The instrument has a pixel size of 0.125 arcsec and reports a spatial resolution of ~0.3 arcsec (*30*). The analyzed quiet sun images were taken with an exposure time of 1 s. The authors report a 'forest of very thin threadlike structures' with lengths around 10 arcsec and with widths in the range 1-3 arcsec.

Based on the more or less uniform brightness of the structures and implying that they do not greatly exceed the gravitational scale height, it is concluded that the structures are low-lying. Since the VAULT observations are not at the limb and do not allow easy identification of "loop"-like structures because of the viewing angle, the authors resort to modeling to deduce the nature of the observed structures. The authors conclude that though their interpretations are suggestive of cool loops, they cannot rule out other interpretations due to the limitations and assumptions inherent in their modeling (see also *31* whom claim that only a small fraction of the emission found in the VAULT observations are due cool loops). In particular, they highlight that time-dependent effects, as well as the effects of flows, need to be incorporated.

One other possible interpretation for the VAULT observations is that the features observed are related to spicules. Indeed, using SUMER spectroheliograms of the O VI 1032 Å line combined with MDI magnetograms Warren & Winbarger (*8*) concluded that the numerous, narrow "looplike" structures they found in their spectroheliogram were due to highly dynamic spicules, not low lying loops. This interpretation highlights the difficulty of separating spicular structures from closed low lying loops in observations taken on the solar disk. As is clear from Movies 1 and 2, UFS loops and spicules are easily separated close to and above the limb given high enough spatial and temporal resolution, such as with the IRIS SJI movies.

Clearly magnetic field lines show collective behavior (heating) on scales that are resolved by the IRIS observations and thus define 'loops' that appear to evolve as a single unit. Note that while we have discussed the UFS as monolithic structures, it is possible that the loops have substructure at scales smaller than that resolved. For example, the numerical simulations show that what are seen as monolithic loops at IRIS resolution display substructure in the simulation's native resolution. Nevertheless, since the process defining the loops has coherence perpendicular to the long axes of the loops on scales larger than that observed, we believe it is useful to study the collective behavior of individual magnetic field strands on these larger scales.

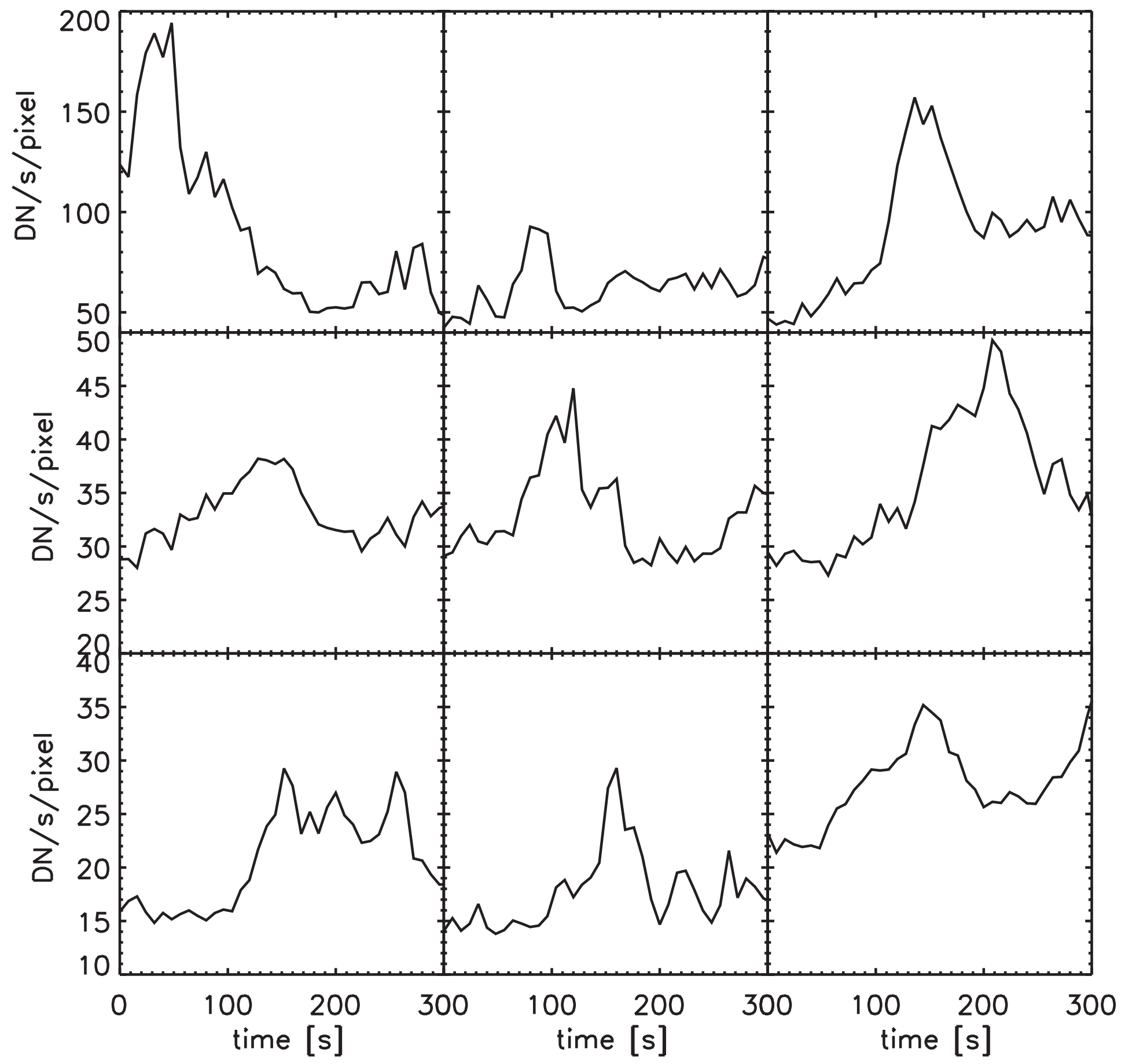

Fig. S1.
Time evolution of selected UFS loops. The shapes of UFS loops picked by inspection from the observations commencing December 9 2013 23:33 UT were fit with fourth order polynomials. The total intensity of the loops is computed by summing the intensity and then dividing by the exposure time (8 s) and number of pixels in the fit.

**Table S1.**

Properties of IRIS datasets used in this paper.

| Date | Type | FOV [arcsec$^2$] | Step, Pixel Size [arcsec] | Exposure time [seconds] | Pointing | Figure/ Movie |
|---|---|---|---|---|---|---|
| 1-Oct-2013 07:51-10:41 | Medium coarse 2-step raster | 2x66 | 2, 0.16 | 4 | East Limb | 3 |
| 2-Oct-2013 07:16-11:14 | Slit Jaw Si IV 1400 movie | 60x60 | 0.16, 0.16 | 4 | North Pole | 2 |
| 9/10-Dec-2013 23:33-00:13 | Slit Jaw Si IV 1400 movie | 120x128 | 0.16, 0.16 | 8 | West Limb | 1/S1,S3 |
| 22-Feb-2014 16:09-17:18 | Large dense 16-step raster | 5 x 128 | 0.35, 0.16 | 8 | North West Limb | S2, S4, S5 |

**Movie S1**

Time evolution of UFS on the western solar limb constructed with slit jaw images taken in the Si IV 1400 Å filter on 9-Dec-2013. Emission far above the limb is slightly enhanced by scaling the images intensity to $I^{\gamma}$ with $\gamma$=0.7. Note that while many UFS loops are short lived, nests of loops occur in which new loops appear to be formed continuously during the entire observation period.

**Movie S2**

As Movie S1, time evolution of UFS on the north west solar limb in the Si IV 1400 Å slit jaw filter on 22-Feb-2014. The intensity off the limb is enhanced by scaling the images intensity to $I^{\gamma}$ with $\gamma$=0.7.

**Movie S3**

Time evolution of UFS on the western solar limb (same as in Movie S1) as observed with the Si IV 1400 Å slit jaw filter on 9-Dec-2013. In the lower panel this is compared with co-spatial and co-temporal AIA observations taken in the He II 304 Å filter in red, the Fe IX/X 171 Å filter in blue, and the Fe XII 193 Å filter in green.

**Movie S4**

UFS as seen in 'realistic' 3D simulations. The top panel shows the synthetic Si IV 1393 Å spectral line as computed from the electron density, temperature, and line of sight velocity from the simulation using CHIANTI (*26*) compiled atomic data. The middle panel shows the total intensity of the Si IV line, the dashed line indicating the location of the synthetic spectra shown in the top panel. The lower panel shows the Dopplergram of the Si IV line formed by subtracting signals in the blue wing of a spectral line from those in the red wing. Both wings are sampled in 10 km/s wide bands centered on 20 km/s. (See Movies S5 and S6 for a comparison with observations).

**Movie S5**

Si IV 1400 Å slit jaw movies (top panel) along with images formed by spectral scans near the solar north west limb from data taken on 22-Feb-2014. The slit jaw movie has a frame to frame cadence of ~18 s, the spectral scans a cadence of roughly 150 s. The intensities in a 10 km/s band centered 38 km/s towards the blue (second panel from the top) and the red (third panel) are subtracted to form a Dopplergram (fourth panel), and summed intensity (fifth panel). The intensity in the line center is shown in the bottom panel.

**Movie S6**

Same as in Movie S5, but showing blue and red wing intensities and Dopplergrams centered on 49 km/s.